\documentclass[aps,pre,twocolumn,groupedaddress,floatfix,longbibliography,superscriptaddress]{revtex4-1}

\usepackage{amsmath}
\usepackage{amssymb}
\usepackage{amsfonts}
\usepackage{graphicx}
\usepackage{bbold}
\usepackage{bm}
\usepackage{bm}
\usepackage{cancel}
\usepackage{times,float}
\usepackage{graphicx}
\usepackage[usenames,dvipsnames,svgnames]{xcolor}
\usepackage{hyperref}
\hypersetup{colorlinks=true, linkcolor=Blue, citecolor=PineGreen,urlcolor=Blue}
\usepackage{multirow}
\usepackage{ulem}
\usepackage{color,epsfig}
\usepackage{bm}
\usepackage{slashed}
\usepackage{hyperref}
\usepackage{subfigure}
\usepackage{feynmf}

\newcommand{\bea}{\begin{eqnarray}}
\newcommand{\eea}{\end{eqnarray}}
\newcommand{\be}{\begin{equation}}
\newcommand{\ee}{\end{equation}}

\renewcommand\vec{\bm}

\renewcommand\Im{\text{Im}}

\newcommand{\nn}{\nonumber}
\newcommand{\ii}{\mathrm{i}}
\newcommand{\tb}[1]{t_B({#1})}
\newcommand{\qB}{eB}
\newcommand{\kt}{k_\perp}

\newcommand{\kp}{k_\parallel}
\newcommand{\gt}{g_\perp}
\newcommand{\gp}{g_\parallel}

\begin{document}

\title{QED Fermions in a noisy magnetic field background: The effective action approach}
\author{Jorge David Casta\~no-Yepes}
\email{jcastano@uc.cl}
\affiliation{Facultad de F\'isica, Pontificia Universidad Cat\'olica de Chile, Vicu\~{n}a Mackenna 4860, Santiago, Chile}
\author{Marcelo Loewe}
\email{mloewe@fis.puc.cl}
\affiliation{Centre for Theoretical and Mathematical Physics, and Department of Physics, University of Cape Town, Rondebosch 7700, South Africa}
\affiliation{Facultad de Ingeniería, Arquitectura y Diseño, Universidad San Sebastián, Santiago, Chile}
\author{Enrique Mu\~noz}
\email{munozt@fis.puc.cl}
\affiliation{Facultad de F\'isica, Pontificia Universidad Cat\'olica de Chile, Vicu\~{n}a Mackenna 4860, Santiago, Chile}
\affiliation{Center for Nanotechnology and Advanced Materials CIEN-UC, Avenida Vicuña Mackenna 4860, Santiago, Chile}
\author{Juan Crist\'obal Rojas}
\email{jurojas@ucn.cl}
\affiliation{Departamento de Física, Universidad Cat\'olica del Norte, Angamos 610, Antofagasta, Chile}

\date{\today}

\begin{abstract}
We consider the effects of a noisy magnetic field background over the fermion propagator in QED, as an approximation to the spatial inhomogeneities and time-fluctuations that would naturally arise in certain physical scenarios, such as heavy-ion collisions or the quark-gluon plasma in the early stages of the evolution of the Universe. We considered a classical, finite and uniform average magnetic field background $\langle \mathbf{B}(\mathbf{x})\rangle_{\Delta} = \mathbf{B}$, subject to white-noise fluctuations with auto-correlation of magnitude $\Delta_B$. By means of the Schwinger representation of the propagator in the average magnetic field as a reference system, we used the replica formalism to study the effects of the magnetic noise at the mean field level, in terms of a vector order parameter $Q_j = \ii e \Delta_B\langle\langle \bar{\psi}\gamma_j\psi \rangle\rangle_{\Delta}$ whose magnitude represents the ensemble average (over magnetic noise) of the fermion currents. We identified the region where this order parameter acquires a finite value, thus breaking the $U(1)$-symmetry of the model due to the presence of the magnetic noise.

\end{abstract}

\maketitle 

\section{Introduction}

There is a number of important physical scenarios where the presence of strong magnetic fields determine the dynamics of relativistic particles. The quark-gluon plasma~\cite{Busza_2018,Hattori_2016,Hattori_2018,Buballa_2005} and heavy-ion collisions~\cite{Alam_2021,Inghirami_2020,Ayala:2019jey,Ayala:2017vex} are among them. For technical reasons, it is a common assumption in the theoretical analysis of such systems to simplify the configuration of the field, by assuming it to be constant (both static and uniform), such that the Schwinger proper time formalism can be applied~\cite{Schwinger_1951,Dittrich_Reuter,Dittrich_Gies}. However, in a more realistic description of such phenomena, the electromagnetic field may develop spatio-temporal patterns~\cite{Inghirami_2020,Alam_2021} that will then in principle modify the associated physical predictions. We discussed such possibility in our recent work \cite{Replicas_1}, where we applied a perturbative analysis to conclude that such magnetic noise effects may indeed be relevant. In this work, we revisit the problem from a non-perturbative point of view, in order to shed further light into such effects over a broader range of magnetic noise intensities.

Following the analysis presented in our previous work~\cite{Replicas_1},
we shall consider a physical scenario where a classical and static magnetic field background, possessing local random fluctuations, modifies the quantum dynamics of a system of fermions. For this purpose,
we shall assume the standard QED theory involving fermionic fields $\psi(x)$, as well as gauge fields $A^{\mu}(x)$. In the later, we shall distinguish three physically different contributions~\cite{Replicas_1}
\begin{eqnarray}
A^{\mu}(x) \rightarrow A^{\mu}(x) + A^{\mu}_\text{BG}(x) + \delta A^{\mu}_\text{BG}(x).
\label{eq_Atot}
\end{eqnarray}

Here, $A^{\mu}(x)$ represents the dynamical photonic quantum field, while BG stands for ``background", thus capturing the presence of a classical external field $\mathbf{B} = \langle\nabla\times\mathbf{A}_\text{BG}(x)\rangle_{\Delta}$, assumed to be static and uniform as imposed by the experimental conditions. Moreover, for this BG contribution, we consider the effect of white noise fluctuations $\delta A^{\mu}_\text{BG}(x)$ with respect to the mean value $A_\text{BG}^{\mu}(x)$, satisfying the statistical properties
\begin{eqnarray}
\langle \delta A^{j}_\text{BG}(x) \delta A^{k}_\text{BG}(x')\rangle_{\Delta} &=&  \Delta_{B}\delta_{j,k}\delta^{(4)}(x-x'),\nonumber\\
\langle \delta A^{\mu}_\text{BG}(x)\rangle_{\Delta} &=& 0.
\label{eq_Acorr}
\end{eqnarray}
We remark that in contrast with our previous work~\cite{Replicas_1}, where we assumed only spatial fluctuations, here we shall assume spatio-temporal fluctuations in the background gauge fields.
These statistical properties are represented by a Gaussian functional distribution of the form
\begin{eqnarray}
dP\left[ \delta A^{\mu}_\text{BG} \right] = \mathcal{N} e^{-\int d^4x\,\frac{\left[\delta A_\text{BG}^{\mu}(x)\right]^2}{2 \Delta_B}}
\mathcal{D}\left[\delta A_\text{BG}^{\mu}(x)\right],
\label{eq_Astat}
\end{eqnarray}
with the corresponding statistical average over background fluctuations defined by
\begin{eqnarray}
\langle \hat{O} \rangle_{\Delta} = \int dP\left[ \delta A^{\mu}_\text{BG} \right] \hat{O}\left[ \delta A^{\mu}_\text{BG}\right].
\label{eq_def_average}
\end{eqnarray}

In heavy-ion collisions (HIC), strong magnetic fields $\mathbf{B} = \nabla\times\mathbf{A}_\text{BG}$ are generated locally within a small spatial region whose characteristic length scale is $L \sim \sqrt{\sigma}$, with $\sigma$ being the effective cross-section. In these collisions, the dominant component of the magnetic field is along the axial z-direction, such that on average we have $\langle \mathbf{B} \rangle = \hat{e}_3\,B$. However, there are also smaller transverse components $\delta B_x$ and $\delta B_y$ such that we can estimate the fluctuation of the field within the small collision region to be on the order of $(\delta B)^2 \sim (\delta B_x)^2 + (\delta B_y)^2$. Since many such collisions occur at different points in space and their time-span $\delta \tau \sim L'/c$, an approximate model for this physical scenario is provided by the magnetic random noise Eq.~\eqref{eq_Acorr}. By dimensional analysis, the magnitude of $\Delta_B$ is of the order:
\begin{equation}
\Delta_B \sim \left(\delta B\right)^2\,L^{5} L' \sim \left(\delta B\right)^2\,\sigma^{5/2} L'.
\label{eq_DeltaB}
\end{equation}

The effective cross-section for a nuclear collision can be estimated as the fraction $f$ of the area of perfectly central collisions between two nuclei, each with a radius of $r_A$
\begin{equation}
\sigma = f\pi r_A^2.
\end{equation}

Here, $f$ represents the fraction of the geometrical cross-section $\sigma_\text{geom}$, which is defined as the area of the circle with a radius of $r_1 + r_2 = 2R$ in a maximum peripheral collision, and the cross-section $\sigma_b$ for a peripheral collision with impact parameter $b$ \cite{bartke2008introduction, castano2021effects}:
\begin{equation}
f = \frac{\sigma_b}{\sigma_\text{geom}} = \left(\frac{N_\text{part}}{2N}\right)^{2/3},
\end{equation}
where $\sigma_b$ describes an effective nucleus of radius $b$. The nuclear radius is always written as $r_A = r_0N^{1/3}$, where $N$ is the number of nucleons per ion and $r_0 \sim 10^{-3}\,\text{MeV}^{-1}$. Here, $N_\text{part}$ is the number of participants corresponding to the effective nucleus.

Therefore, under these considerations, we have
\begin{equation}
\Delta_B \sim \pi^{5/2}\left(\delta B\right)^2r_0^5N^{5/3}\left(\frac{N_\text{part}}{2N}\right)^{5/3}L'.
\end{equation}

In peripheral heavy-ion collisions, the magnetic field fluctuations along the transverse plane are approximately $|e\,\delta B| \sim m_\pi^2/4$, where $m_\pi$ is the pion mass \cite{PhysRevC.83.054911, castano2021effects}. For an Au+Au collision with $N=197$, and if $N_\text{part}/N=1/2$, we obtain (for $L' \sim r_A$)
\begin{equation}
e^2\Delta_B \sim 1.6\times 10^{-6}\,\text{MeV}^{-2},
\end{equation}
or for less central collisions with $N_\text{part}/N=1/8$:
\begin{equation}
e^2\Delta_B \sim 1.6\times10^{-7}\,\text{ MeV}^{-2}.
\end{equation}

As we shall later show, the effects of magnetic noise are controlled by the dimensionless parameter $\Delta = e^2\Delta_B m_f^2$, where $m_f$ is the fermion mass. If one considers the mass of the proton, then the relevant dimensionless scale would be
\begin{eqnarray}
\Delta_{\text{proton}} \sim 0.16 - 1.6,
\end{eqnarray}
 whereas if one considers the mass of the constituent quark species

\begin{eqnarray}
\Delta_{\text{quark}} \sim 0.018 - 0.18.
\end{eqnarray}

Taking into account the previous analysis, we write the Lagrangian for this model as a superposition of two terms

\begin{eqnarray}
\mathcal{L} = \mathcal{L}_\text{FBG} + \mathcal{L}_\text{NBG}.
\label{eq_LAG}
\end{eqnarray}

Here, the first term represents the system of Fermions (and photons) immersed in the deterministic background field (FBG)
\begin{eqnarray}
\mathcal{L}_\text{FBG} = \bar{\psi}\left(\ii\slashed{\partial} - e \slashed{A}_\text{BG} - e \slashed{A}   - m_f \right)\psi-\frac{1}{4}F_{\mu\nu}F^{\mu\nu},
\label{eq_LFBG}
\end{eqnarray}
where $F_{\mu\nu} = \partial_{\mu} A_{\nu}- \partial_{\nu} A_{\mu}$ is the strength tensor for the dynamical quantum gauge fields (photons). The second term in the lagrangian Eq.~\eqref{eq_LAG} represents the interaction between the Fermions and the classical noise (NBG)
\begin{eqnarray}
\mathcal{L}_\text{NBG} = \bar{\psi}\left( - e \delta\slashed{A}_\text{BG} \right)\psi.
\label{eq_LDBG}
\end{eqnarray}

The generating functional (in the absence of sources) for a given realization of the noisy  fields is given by
\begin{eqnarray}
Z[A] = \int \mathcal{D}[\bar{\psi},\psi]
e^{\ii\int d^4 x \left[ \mathcal{L}_\text{FBG} + \mathcal{L}_\text{NBG}  \right]}.
\end{eqnarray}

To study the physics of this system, as usual we need the generator of connected correlation functions. However, as the presence of disorder is modeled by a statistical ensemble of different realizations of the magnetic background noise $\delta A_\text{BG}^{\mu}(x)$, we need to calculate the disorder-averaged generator of connected correlation functions $\langle \ln Z \rangle_{\Delta}$. For this purpose, we apply the replica method, which is based on the following identity~\cite{Parisi_Mezard_1991}
\begin{eqnarray}
\langle \ln Z[A]\rangle_{\Delta} = \lim_{n\rightarrow 0}\frac{\langle Z^n[A]\rangle_{\Delta}-1}{n}.
\label{eq_ln}
\end{eqnarray}

Here, we defined the statistical average according to the Gaussian functional measure of Eq.~\eqref{eq_Astat} as in Eq.~\eqref{eq_def_average}, and $Z^n$ is obtained by incorporating an additional ``replica" component for each of the Fermion fields, i.e. $\psi(x)\rightarrow \psi^{a}(x)$, for $1\le a \le n$. The ``replicated" Lagrangian has the same form as Eqs.~\eqref{eq_LFBG} and \eqref{eq_LDBG}, but with an additional sum over the replica components of the Fermion fields. Therefore, the averaging procedure leads to
\begin{eqnarray}
\langle Z^n[A] \rangle_{\Delta} &=& \int 
\prod_{a=1}^{n}\mathcal{D}[\bar{\psi}^{a},\psi^{a}]
\int \mathcal{D}\left[\delta A_\text{BG}^{\mu}\right]e^{-\int d^4x\,\frac{\left[\delta A_\text{BG}^{\mu}(x)\right]^2}{2 \Delta_B}}\nonumber\\
&&\times e^{\ii\int d^4 x \sum_{a=1}^n \left( \mathcal{L}_\text{FBG}[\bar{\psi}^a,\psi^a] + \mathcal{L}_{DBG}[\bar{\psi}^a,\psi^a] \right)}\nonumber\\
&=& \int 
\prod_{a=1}^{n}\mathcal{D}[\bar{\psi}^{a},\psi^{a}] e^{\ii \bar{S}\left[\bar{\psi}^a,\psi^a;A \right]},
\label{eq_repl}
\end{eqnarray}
where in the last step we explicitly performed the Gaussian integral over the background noise, leading to the definition of the effective averaged action for the replica system
\begin{widetext}
\begin{eqnarray}
\ii\,\bar{S}\left[\bar{\psi}^a,\psi^a;A \right]
&=& \ii\,\int d^4 x \left(\sum_{a}\bar{\psi}^{a}\left(\ii\slashed\partial -  e \slashed{A}_\text{BG} - e \slashed{A} - m_f  \right)\psi^{a}-\frac{1}{4}F_{\mu\nu}F^{\mu\nu}\right)\nonumber\\
&-& \frac{e^2\Delta_{B}}{2}\int d^4x\int d^4 y\sum_{a,b}\sum_{j=1}^{3}\bar{\psi}^{a}(x)\gamma^{j}\psi^{a}(x)\bar{\psi}^{b}(y)\gamma_{j}\psi^{b}(y).
\label{eq_Savg}
\end{eqnarray}
\end{widetext}

Clearly, we end up with an effective interacting theory between vector currents corresponding to different replicas, with a coupling constant proportional to the fluctuation amplitude $\Delta_B$ that characterizes the magnetic noise, as defined in Eq.~\eqref{eq_Acorr}. In this context, the noise parameter plays the role of a mutual induction coefficient between such currents. 

The ``free'' part of the action corresponds to Fermions in the average background classical field
$A_\text{BG}^{\mu}(x)$. We choose this background to represent a uniform, static magnetic field along the $z$-direction $\mathbf{B} = \hat{e}_3 B$, using the gauge~\cite{Dittrich_Reuter}
\begin{eqnarray}
A_\text{BG}^{\mu}(x) = \frac{1}{2}(0,-B x^2, B x^1,0).
\label{eq_BGauge}
\end{eqnarray}

In addition, as we shall focus on the analysis for the fermion propagator, we shall not consider the photons in this scenario $A^{\mu}(x) = 0$, and hence its corresponding strength tensor $F_{\mu\nu} = 0$.

\section{Introduction of auxiliary bosonic fields}
Let us now introduce a Hubbard-Stratonovich transformation via a set of complex bosonic fields $Q_{j}(x)$, by means of the Gaussian integral identity
\begin{widetext}
\begin{eqnarray}
&&e^{ -\frac{e^2\Delta_B}{2}\int d^4 x\int d^4 y \sum_{a,b}^{n}\sum_{j=1}^{3}\bar{\psi}^{a}(x)\gamma^{j}\psi^{a}(x)\bar{\psi}^{b}(y)\gamma_{j}\psi^{b}(y) }\nonumber\\
&&= \mathcal{N} \left[\prod_{j=1}^{3}\int\mathcal{D}Q_j (x)\mathcal{D}Q_j^{*} (x) \right]
 e^{-\frac{2}{\Delta_{B}} \int d^4 x \left|Q_{j}(x)\right|^2 + \ii e \int d^4 x Q_{j}(x)\sum_{a=1}^n \bar{\psi}^{a}(x)\gamma^{j}\psi^{a}(x)- \ii e \int d^4 x Q_{j}^{*}(x)\sum_{a=1}^n \bar{\psi}^{a}(x)\gamma^{j}\psi^{a}(x)}\nonumber\\ 
\end{eqnarray}

With this transformation into the averaged, replicated $n^{th}$ power of the generating functional Eq.~\eqref{eq_repl}, we obtain the equivalent form

\begin{eqnarray}
\langle Z^n[A_\text{BG}]\rangle_{\Delta} &=& \mathcal{N} \left[\prod_{j=1}^{3}\int\mathcal{D}Q_j (x)\mathcal{D}Q_j^{*} (x)\right]
e^{-\frac{2}{\Delta_{B}} \int d^4 x\, \left|Q_{j}(x)\right|^2  }\nonumber\\
&&\times
\left[\prod_{a=1}^{n}\int\mathcal{D}[\bar{\psi}^{a},\psi^{a}]\right] e^{\ii\int d^4 x \left\{\sum_{a=1}^{n}\bar{\psi}^{a}(x)\left(\ii\slashed\partial -  e \slashed{A}_\text{BG} - m_f  - e \gamma^{j}(Q_j - Q_j^*)\right)\psi^{a}(x)\right\}  }\nonumber\\
&=& \mathcal{N} \left[\prod_{j=1}^{3}\int\mathcal{D}Q_j (x)\mathcal{D}Q_j^{*} (x) \right]
e^{-\frac{2}{\Delta_{B}} \int d^4 x\, \left|Q_{j}(x)\right|^2  }
\left[\det\left( \ii\slashed\partial -  e \slashed{A}_\text{BG} - m_f   - e \gamma^{j}\left( Q_j - Q_j^* \right) \right)  \right]^{n}\nonumber\\
&=& \mathcal{N} \left[\prod_{j=1}^{3}\int\mathcal{D}Q_j (x)\mathcal{D}Q_j^{*} (x) \right]
e^{-\frac{2}{\Delta_{B}} \int d^4 x\, \left|Q_{j}(x)\right|^2  + n{\rm{Tr}}\ln\left[ \ii\slashed\partial -  e \slashed{A}_\text{BG} - m_f  - e \gamma^{j}\left( Q_j - Q_j^* \right)\right]  }
\label{eq_barZn}
\end{eqnarray}

Based on this identity, and combined with Eq.~\eqref{eq_ln}, we obtain the effective action
\begin{eqnarray}
&&\ii\,S_\text{eff}[A_\text{BG}] - \ii\,S_{0}[A_\text{BG}]  = \langle \ln Z[A_\text{BG}]\rangle_{\Delta} - \ln Z_0[A_\text{BG}]\nonumber\\
&=&  \mathcal{N} \left[\prod_{j=1}^{3}\int\mathcal{D}Q_j (x)\mathcal{D}Q_j^{*} (x)  \right]
e^{-\frac{2}{\Delta_{B}} \int d^4 x\, \left|Q_{j}(x)\right|^2} \lim_{n\rightarrow 0}\frac{1}{n}\left[e^{n{\rm{Tr}}\ln\left[ \ii\slashed\partial -  e \slashed{A}_\text{BG}  - m_f - e\gamma^{j}(Q_j-Q_j^*) \right]  }-1\right]- \ln Z_0[A_\text{BG}]\nonumber\\
&=& \langle {\rm{Tr}} \ln\left( \ii\slashed\partial -  e \slashed{A}_\text{BG} - m_f  - e \gamma^{j}(Q_j - Q_j^*) \right) \rangle_{\Delta} - {\rm{Tr}} \ln\left( \ii\slashed\partial -  e \slashed{A}_\text{BG} - m_f  \right),
\label{eq_effact}
\end{eqnarray}
\end{widetext}
where $\langle (\cdot) \rangle_{\Delta}$ represents the average over the Gaussian functional measure of the complex fields $Q_{j}(x)$. 
Let us define the inverse fermion propagator, including the classical background field, as follows
\begin{eqnarray}
S_\text{F}^{-1}(x-y) = \left( \ii\slashed{\partial} - e \slashed{A}_\text{BG} - m_f \right)_{x}\delta^{(4)}(x-y)
\label{eq_Schwingernonoise}
\end{eqnarray}

Therefore, for the effective action we have
\begin{eqnarray}
\ii\,S_\text{eff}[A_\text{BG}] &=& \langle {\rm{Tr}}\ln\left( \ii\slashed{\partial} - e\slashed{A}_\text{BG} - m_f - e \gamma^{j}(Q_j - Q_j^*)  \right) \rangle_{\Delta}\nonumber\\
&=& \langle {\rm{Tr}}\ln\left(  S_\text{F}^{-1} - e \gamma^{j}(Q_j - Q_j^*) \right) \rangle_{\Delta}\nonumber\\
&=& {\rm{Tr}}\ln S_\text{F}^{-1}\nonumber\\ 
&+& 
\langle {\rm{Tr}}\ln\left( \mathbf{1} - e S_\text{F} \gamma^{j}(Q_j - Q_j^*)   \right) \rangle_{\Delta}.
\end{eqnarray}

By noticing that $\ii\,S_{0}[A_\text{BG}] = {\rm{Tr}}\ln S_\text{F}^{-1}$, we have
\begin{eqnarray}
\ii\,S_\text{eff}[A_\text{BG}] &-& \ii\,S_{0}[A_\text{BG}]\nonumber\\ &=& \langle {\rm{Tr}}\ln\left( \mathbf{1} - e S_\text{F} \gamma^{j}(Q_j - Q_j^*)  \right) \rangle_{\Delta}\nonumber\\
\label{eq_deltaS}
\end{eqnarray}
where the right hand side contains all the effects of the noise.
\section{Saddle-point approximation (Mean-field)}
The result in Eq.~\eqref{eq_deltaS} can be expressed in the explicit functional integral form
\begin{widetext}
\begin{eqnarray}
\ii\,S_\text{eff}[A_\text{BG}] - \ii\,S_{0}[A_\text{BG}] = \mathcal{N} \left[\prod_{j=1}^{3}\int\mathcal{D}Q_j (x)\mathcal{D}Q_j^{*} (x)  \right]
e^{-\frac{2}{\Delta_{B}} \int d^4 x\, \left|Q_{j}(x)\right|^2 + \ln\left[{\rm{Tr}}\ln\left( \mathbf{1} - e S_\text{F} \gamma^{j}(Q_j - Q_j^*)  \right)\right]}.
\label{eq_effS_saddle}
\end{eqnarray}

In order to study the effects of the background noise, we shall adopt a mean-field approximation, by searching for the saddle-point of the exponent in Eq.~\eqref{eq_barZn}

\begin{eqnarray}
\frac{\delta}{\delta Q_j(x)}\left\{-\frac{2}{\Delta_B}\int d^4 y Q_{l}(y)Q_{l}^{*}(y) + \ln\left[{\rm{Tr}}\ln\left[ \mathbf{1}- e S_\text{F} \gamma^{l}(Q_l - Q_l^*)  \right]\right]\right\} &=& 0,\nonumber\\
\frac{\delta}{\delta Q_j^{*}(x)}\left\{-\frac{2}{\Delta_B}\int d^4 y Q_{l}(y)Q_{l}^{*}(y)+ \ln\left[{\rm{Tr}}\ln\left[ \mathbf{1}- e S_\text{F} \gamma^{l}(Q_l - Q_l^*)  \right]\right]\right\} &=& 0.
\label{eq_variational}
\end{eqnarray}
\end{widetext}

This condition leads to the equation (assuming homogeneous solutions of the form $Q_j(x)\equiv Q_j$)
\begin{eqnarray}
Q_j^{*} &=& -e\frac{\Delta_B}{2} \frac{{\rm{Tr}}\left[S_\text{F}\gamma^{j}\left( \mathbf{1}- e S_\text{F} \gamma^{l}(Q_l - Q_l^*)   \right)^{-1}\right]}{{\rm{Tr}}\ln\left[ \mathbf{1}- e S_\text{F} \gamma^{l}(Q_l - Q_l^*)  \right]}
\nonumber\\
\eea
From the second equation in Eq.~\eqref{eq_variational}, we obtain the additional condition
\bea
Q_j^{*} = - Q_j.
\eea
We can combine both equations, by noticing that
\bea
Q_j + Q_j^{*} &=& 0\nonumber\\
q_j \equiv Q_j - Q_j^{*} &=& 2\, \ii\, \Im Q_j\ne 0,
\label{eq_qj}
\eea
where the second line implies that $q_j$ is a pure imaginary number, and it satisfies the non-linear equation
\bea
q_j &=& e\Delta_B \frac{{\rm{Tr}}\left[S_\text{F}\gamma^{j}\left( \mathbf{1}- e S_\text{F} \gamma^{l}q_l   \right)^{-1}\right]}{{\rm{Tr}}\ln\left[ \mathbf{1}- e S_\text{F} \gamma^{l}q_l  \right]}.
\label{eq_MFexact}
\eea

The numerator of this equation can be expanded as an infinite geometric series as follows
\begin{eqnarray}
&&{\rm{Tr}}\left[S_\text{F}\gamma^{j}\left( \mathbf{1}- e S_\text{F} \gamma^{l}q_l   \right)^{-1}\right] = \sum_{l=0}^{\infty}e^{l}{\rm{Tr}} \left[ S_\text{F}\gamma^{j}\left(  S_\text{F}\gamma^{k}q_k\right)^{l} \right]\nonumber\\
&&=  \sum_{l=1}^{\infty}e^{l}\left[\Pi_{\alpha=1}^{l}q_{k_{\alpha}} \right]{\rm{Tr}} \left[ S_\text{F}\gamma^{j}  S_\text{F}\gamma^{k_{1}}S_\text{F}\gamma^{k_{2}}\ldots S_\text{F}\gamma^{k_{l}}  \right].
\end{eqnarray}

On the other hand, the denominator can also be expanded by means of the Taylor series for the natural logarithm, as follows
\bea
&&{\rm{Tr}}\ln\left[ \mathbf{1}- e S_\text{F} \gamma^{l}q_l  \right]
= \sum_{l=1}^{\infty}\frac{e^l}{l}{\rm{Tr}}\left[ \left(S_F \slashed{q}\right)^{l} \right]\nn\\
&=& \sum_{l=1}^{\infty}\frac{e^l}{l}\left[\Pi_{\alpha=1}^{l}q_{k_{\alpha}} \right]{\rm{Tr}}\left[   S_\text{F}\gamma^{k_{1}}S_\text{F}\gamma^{k_{2}}\ldots S_\text{F}\gamma^{k_{l}} \right].
\eea

From the exact expression Eq.~\eqref{eq_MFexact}, we can extract the leading contribution by retaining terms up to third order in the $q_j$ coefficients in the numerator, while retaining up to second order in the denominator
\begin{eqnarray}
q_j = e\Delta_B \frac{e\mathcal{M}^{jl}q_l + e^3\mathcal{M}^{jlmn}q_l q_m q_n}{\frac{e^2}{2}\mathcal{M}^{mn}q_m q_n},
\label{eq_system}
\end{eqnarray}
where we defined the matrix coefficients
\begin{eqnarray}
\mathcal{M}^{jl} &=& {\rm{Tr}} \left[ S_\text{F}\gamma^{j}S_\text{F}\gamma^{l}\right]\nn\\
&=& \int \frac{d^4 k}{(2\pi)^4} {\rm{tr}} \left[ S_\text{F}(k)\gamma^{j}S_\text{F}(-k)\gamma^{l}\right],
\label{m2}
\end{eqnarray}
and
\begin{eqnarray}
&&\mathcal{M}^{jlmn} = {\rm{Tr}} \left[ S_\text{F}\gamma^{j}S_\text{F}\gamma^{l}S_\text{F}\gamma^{m}S_\text{F}\gamma^{n}\right]
\label{m4}
\\
&=& \int \frac{d^4 k}{(2\pi)^4} {\rm{tr}}\left[ S_\text{F}(k)\gamma^{j}S_\text{F}(k)\gamma^{l}S_\text{F}(k)\gamma^{m}S_\text{F}(k)\gamma^{n}\right].\nn
\end{eqnarray}

Here, ${\rm{tr}}[ \cdot ]$ stands for trace over the space of Dirac matrices. The Schwinger propagator in Fourier space is defined by Eq.~\eqref{eq_Sprop} and, more importantly for calculation purposes, by its alternative form Eq.~\eqref{propSchwinger}.

We can analyze the possible solutions to Eq.~\eqref{eq_system}, by casting it into the form of a quasi-linear system
\begin{eqnarray}
\left( \Delta_B \mathbf{\mathcal{M}} + \tilde{\mathbf{\mathcal{M}}}[\mathbf{q}] \right)\mathbf{q} = 0,
\label{eq_linear}
\end{eqnarray}
where we defined
\begin{eqnarray}
\left[\tilde{\mathbf{\mathcal{M}}}[\mathbf{q}]\right]^{jl} \equiv \left( -\frac{1}{2}\delta^{jl}\mathcal{M}^{mn} + e^2\Delta_B \mathcal{M}^{jlmn} \right)q_m q_n.
\label{eq_Mtilde_gen}
\end{eqnarray}

There is always a trivial solution $\mathbf{q} = 0$ to Eq.~\eqref{eq_linear}. However, nontrivial solutions $\mathbf{q}$ may exist provided that the (nonlinear) matrix coefficient is singular, i.e.
\begin{eqnarray}
\det \left( \Delta_B \mathbf{\mathcal{M}} + \tilde{\mathbf{\mathcal{M}}}[\mathbf{q}] \right) = 0.
\label{eq_secular}
\end{eqnarray}

In order to analyze this second condition, we need to evaluate the matrix coefficients explicitly. For this purpose, we first discuss the mathematical representation of the Schwinger propagator in the next section.

\begin{figure*}
    \centering
    \includegraphics[width=1\textwidth]{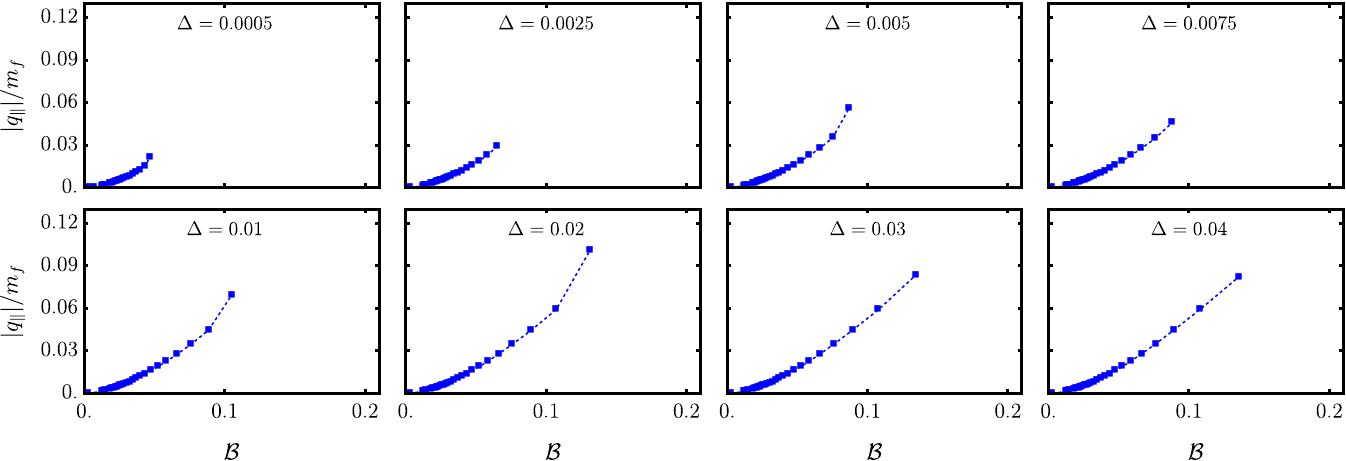}
    \caption{Non-trivial solutions of Eq.~\eqref{eq:det1} for the Case 1 as a function of $\mathcal{B}$ for $\Delta\in[0.1,0.04]$. The dashed line represents the smooth envelope connecting the discrete non-trivial solutions.}
    \label{fig:q3VseB1}
\end{figure*}

\begin{figure*}
    \centering
    \includegraphics[width=1\textwidth]{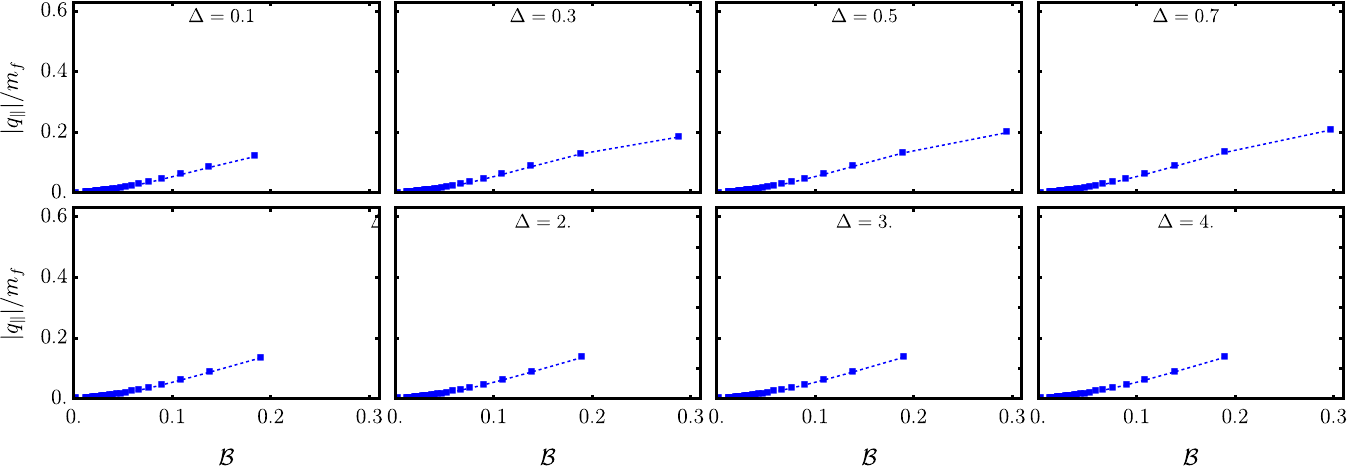}
    \caption{Non-trivial solutions for $q_3$ of Eq.~\eqref{eq:det1} as a function of $eB$ for $\Delta_B\in[0.1,4]$.}
    \label{fig:q3VseB2}
\end{figure*}

\section{The Schwinger propagator}

As discussed in the previous section, the matrix coefficients depend on traces and integrals involving products of the fermion propagator immersed in the constant background magnetic field.
Therefore, this allows us to use directly the Schwinger proper-time representation of the free-Fermion propagator dressed by the background field, whose direction is chosen along the $z$-axis, $\mathbf{B} = \hat{e}_3\,B$, as follows~\cite{Schwinger_1951,Dittrich_Reuter}
\begin{eqnarray}
\left[S_\text{F}(k)\right]_{a,b}
&&= -\ii\delta_{a,b}\int_{0}^{\infty}\frac{d\tau}{\cos(\qB \tau)}
e^{\ii\tau\left(k_{\parallel}^2 - \mathbf{k}_{\perp}^2\frac{\tan(\qB\tau)}{\qB\tau}-m^2_f + \ii\epsilon \right)}\nonumber\\
&&\times\left\{
\left[\cos(\qB\tau) + \gamma^1\gamma^2\sin(\qB\tau)  \right](m_f + \slashed{k}_{\parallel})\right.\nonumber\\
&&\left.+\frac{\slashed{k}_{\perp}}{\cos(\qB \tau)}
\right\},
\label{eq_Sprop}
\end{eqnarray}
which is clearly diagonal in replica space. Here, as usual, we separated the parallel from the perpendicular directions with respect to the background external magnetic field by splitting the metric tensor as $g^{\mu\nu} = g_{\parallel}^{\mu\nu} + g_{\perp}^{\mu\nu}$, with
\begin{eqnarray}
g_{\parallel}^{\mu\nu} &=& \text{diag}(1,0,0,-1),\nonumber\\
g_{\perp}^{\mu\nu} &=& \text{diag}(0,-1,-1,0),
\end{eqnarray}
thus implying that for any 4-vector, such as the momentum $k^{\mu}$, we write
\begin{eqnarray}
\slashed{k} = \slashed{k}_{\perp} + \slashed{k}_{\parallel},
\end{eqnarray}
and
\begin{eqnarray}
k^2 = k_{\parallel}^2 - \mathbf{k}_{\perp}^2,
\end{eqnarray}
respectively. In particular, we have $k_{\parallel}^2 = k_0^2 - k_3^2$,
while $\mathbf{k}_{\perp}=(k^1,k^2)$ is the Euclidean 2-vector lying in the plane perpendicular to the field, such that its square-norm is  $\mathbf{k}_{\perp}^2 = k_1^2 + k_2^2$.
The Schwinger propagator can alternatively be expressed as~\cite{Replicas_1}
\begin{eqnarray}
&&\left[S_\text{F}(k) \right]_{a,b} = -\ii\delta_{a,b}
\left[
\left( m_f + \slashed{k} \right)\mathcal{A}_{1}\right.\nn\\
&&\left.
+ (\ii \qB) \gamma^{1}\gamma^{2}\left( m_f + \slashed{k}_{\parallel} \right)\frac{\partial\mathcal{A}_1}{\partial \mathbf{k}_{\perp}^2} + \left(\ii \qB \right)^2\slashed{k}_{\perp}\frac{\partial^2\mathcal{A}_1}{\partial (\mathbf{k}_{\perp}^2)^2}
\right]\nonumber\\
&&=-\ii\delta_{a,b}\left[ 
\left( m_f + \slashed{k}_{\parallel} \right)\mathcal{A}_1
+  \gamma^{1}\gamma^{2}\left( m_f + \slashed{k}_{\parallel} \right)  \mathcal{A}_2
+ \mathcal{A}_3 \slashed{k}_{\perp}
\right]\nonumber\\
\label{propSchwinger}
\end{eqnarray}

Here, we defined the function
\begin{eqnarray}
\mathcal{A}_1(k,B) = \int_{0}^{\infty}d\tau e^{\ii\tau\left( k_{\parallel}^2 - m_f^2 + \ii\epsilon\right) -\ii\frac{\mathbf{k}_{\perp}^2}{\qB}\tan(\qB \tau) },
\label{eq:A1}
\end{eqnarray}
that clearly reproduces the scalar propagator (with Feynman prescription) in the zero-field limit
\begin{eqnarray}
\lim_{B\rightarrow 0}\mathcal{A}_1(k,B) = \frac{\ii}{k^2 - m_f^2 + \ii\epsilon}, 
\end{eqnarray}
and its derivatives
\begin{subequations}
\bea
\mathcal{A}_2(k,B)&\equiv&\int_0^\infty d\tau ~\tan(\qB\tau)e^{\ii\tau\left(k_\parallel^2-\tb{\tau}\mathbf{k}_\perp^2-m^2_f+\ii\epsilon\right)}\nn\\
&=&\ii \qB\frac{\partial\mathcal{A}_1}{\partial(\mathbf{k}_{\perp}^2)},
\eea
\bea
\mathcal{A}_3(k,B)&\equiv&\int_0^\infty \frac{d\tau}{\cos^2(\qB\tau)}e^{\ii\tau\left(k_\parallel^2-\tb{\tau}\mathbf{k}_\perp^2-m_f^2+\ii\epsilon\right)}\nn\\
&=&\mathcal{A}_1+(\ii \qB)^2\frac{\partial^2\mathcal{A}_1}{\partial(\mathbf{k}_{\perp}^2)^2}.
\eea
\label{A2A3properties}
\end{subequations}

As we showed in detail in our previous work~\cite{Replicas_1}, an exact representation for the function \eqref{eq:A1} is given by
\begin{eqnarray}
\mathcal{A}_1 (k) &=& \frac{\ii\,e^{-\frac{k_{\perp}^2}{eB}}}{2 e B} e^{-\frac{\ii\pi}{2 e B}\left( k_{\parallel}^2 - m_f^2 + \ii\epsilon  \right)}\Gamma\left(-\frac{k_{\parallel}^2 - m_f^2 + \ii\epsilon}{2 e B}  \right)\nonumber\\ &&\times U\left( -\frac{k_{\parallel}^2 - m_f^2 + \ii\epsilon}{2 e B},0,\frac{2k_{\perp}^2}{e B} \right),
\label{eq:A1_hyper}
\end{eqnarray}
where $\Gamma(z)$ is the Gamma function, while $U(a,b,z)$ represents the Tricomi's Confluent Hypergeometric function.

%

\section{Results and Discussion}
\begin{figure*}
    \centering
    \includegraphics[scale=0.8]{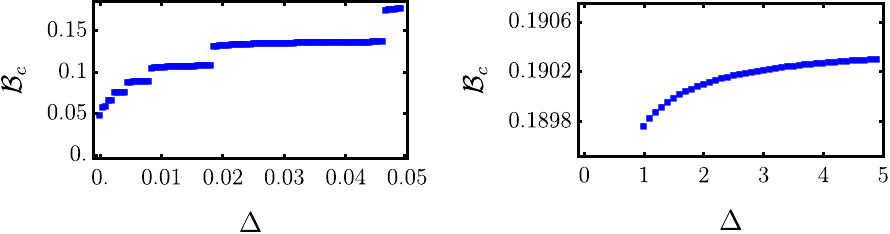}
    \caption{Critical magnetic field $\mathcal{B}_c$ of Fig.~\ref{fig:q3VseB2} as a function of $\Delta$, in two different ranges of such parameter.}
    \label{fig:Bc_vs_Delta_high_1}
\end{figure*}

For the analysis of our numerical results, it is convenient to define the following dimensionless groups,
\bea
\Delta &=& e^2 \Delta_B m_f^2,\nn\\
\mathcal{B} &=& \frac{e B}{m_f^2},
\eea
respectively. 
The matrices $\mathcal{M}^{ij}$ and $\mathcal{M}^{ijkl}$, as defined by Eq.~\eqref{m2} and Eq.~\eqref{m4} respectively, are calculated by tracing over the Dirac matrix space, as explained in Appendix A, and the resulting momentum integrals are calculated from the generic formula developed in Appendix B. The interested reader is referred to those Appendices for further mathematical details.

In order to analyze the existence and features of non-trivial solutions for the order parameter $\vec{q} = (q_1,q_2,q_3)$,
we solve the secular equation Eq.~\eqref{eq_secular}, by assuming two different symmetry conditions, according to the directions orthogonal ($\perp$) and parallel ($\parallel$) to the magnetic field, respectively.

\subsection{ Case 1: $q_3^2 \equiv q_{\parallel}^2$, with $q_1 = q_2 = 0$}

If we impose the condition $q_1=q_2=0$ into Eq.~\eqref{eq_Mtilde_gen}, we have
\begin{eqnarray}
\left[\tilde{\mathcal{M}}[q_{\parallel}]\right]^{jl}
&=&\left( -\frac{1}{2}\delta^{jl}\mathcal{M}^{33} + e^2\Delta_B \mathcal{M}^{jl33} \right)q_{\parallel}^2\nonumber\\
&\equiv& \mathcal{C}^{jl}_{\parallel}q_{\parallel}^2.
\label{eq_case1}
\end{eqnarray}
Therefore, substituting this reduced expression into the secular equation Eq.~\eqref{eq_secular}, we obtain
\bea
{\rm{det}}\left( \Delta_B \mathcal{M} + q_{\parallel}^2 \mathcal{C}_{\parallel}   \right) = 0.
\label{eq:det1}
\eea

Furthermore, using elementary matrix properties and given that the matrix $\mathcal{M}$ is non-singular, we can manipulate the expression above as follows
\bea
&&{\rm{det}}\left( \Delta_B \mathcal{M} + q_{\parallel}^2 \mathcal{C}_{\parallel}   \right) =\\\nonumber 
&&{\rm{det}}\left( \Delta_B \mathcal{M} \right) \cdot {\rm{det}}\left( \mathbf{1}_3 + q_{\parallel}^2 \Delta_B^{-1} \mathcal{M}^{-1} \mathcal{C}_{\parallel}   \right) = 0.
\eea

Given that $\mathcal{M}$ is non-singular, the above expression implies the secular condition
\bea
{\rm{det}}\left( \mathbf{1}_3 + q_{\parallel}^2 \Delta_B^{-1} \mathcal{M}^{-1} \mathcal{C}_{\parallel}   \right) = 0.
\label{eq_sec_case1}
\eea

Our analysis to this point is consistent up to second order powers in the components of the order parameter. Therefore, applying the elementary identity ${\rm{det}}\left(\mathbf{1} + \epsilon X  \right) = 1 + \epsilon {\rm{tr}}X + O(\epsilon^2)$, we expand Eq.~\eqref{eq_sec_case1} to obtain
\bea
q_{\parallel}^2 &=& -\frac{\Delta_B}{{\rm{tr}}\left( \mathcal{M}^{-1} \mathcal{C}_{\parallel}  \right)}\nonumber\\ 
&=& -\frac{\Delta_B}{-\frac{1}{2}\mathcal{M}^{33}{\rm{tr}}\left( \mathcal{M}^{-1}\right) + e^2\Delta_B\rm{tr}\left( \mathcal{M}^{-1} \mathcal{\tilde{M}}^{(33)}  \right)}\nonumber\\
&=& \frac{\chi_{\parallel}\Delta_B}{1 + \mathcal{K}_{\parallel}\Delta_B},
\label{eq:Sigmoid_parallel}
\eea
where we defined the parameters
\bea
\chi_{\parallel} &=& \frac{2}{\mathcal{M}^{33}\,{\rm{tr}}\left( \mathcal{M}^{-1}\right)}\nonumber\\
\mathcal{K}_{\parallel} &=& \frac{-2e^2\rm{tr}\left( \mathcal{M}^{-1} \mathcal{\tilde{M}}^{(33)}  \right)}{\mathcal{M}^{33}\,{\rm{tr}}\left( \mathcal{M}^{-1}\right)},
\eea
as well as the reduced matrix
\bea
\left[\tilde{\mathcal{M}}^{(33)}\right]^{jl} \equiv \mathcal{M}^{jl33}.
\eea

Figure~\ref{fig:q3VseB1} illustrates the non-trivial solutions of Eq.~\eqref{eq:det1} for Case 1, as a function of the external magnetic field and various values of $\Delta$. It can be observed that there exists a region where the discrete solutions exhibit a monotonically increasing pattern with a smooth envelope, abruptly terminated at a point beyond which only the trivial solution $q_\parallel=0$ exists. We refer to this point as the {\it critical magnetic field} $\mathcal{B}_c$. A similar scenario arises when the magnitude of $\Delta$ is 
increased, as demonstrated in Fig.~\ref{fig:q3VseB2}. Furthermore, it is worth noting that for higher values of $\Delta$, the solutions become approximately identical, and the $\mathcal{B}_c$ converges toward a specific limit as is shown in Fig.~\ref{fig:Bc_vs_Delta_high_1}.


In line with the results presented above, it is worth emphasizing that a specific combination of parameters $(\Delta,\mathcal{B})$ plays a pivotal role in giving rise to a discrete set of non-trivial solutions characterized by $q_\parallel\neq 0$, or in the context of Eq.~\eqref{eq_qj}, resulting in purely imaginary values for $q_j$. This behavior is depicted in Figure~\ref{fig:q3VsDelta}, offering valuable insights into the system's dynamics. Figure~\ref{fig:q3VsDelta} illustrates a spectrum of these parameters, revealing that for fixed values of $\mathcal{B}$ not all the values of $\Delta$ produce non-trivial solutions (those are discrete). We have also shown in Figure~\ref{fig:q3VsDelta}, by a dashed line, the smooth envelope of those discrete points.
Nevertheless, for higher values of $\Delta$ the solutions become a quasi-continuum and saturate to the value $q_\parallel^2(\infty)$ defined as
\bea
q_\parallel^2(\infty)=\lim_{\Delta\to\infty}q_\parallel^2=\frac{\chi_\parallel}{\mathcal{K}_\parallel},
\eea
where $q_\parallel^2$ is given in Eq.~\eqref{eq:Sigmoid_parallel}.
In the opposite limit, for very small values of $\Delta$, Eq.~\eqref{eq:Sigmoid_parallel} shows that the order parameter follows an approximately linear trend with a slope defined by
\bea
\chi_{\parallel} = \lim_{\Delta_B\rightarrow 0}\frac{\partial}{\partial \Delta_B} q_{\parallel}^2.
\eea

\begin{figure}
    \centering
    \includegraphics[width=0.5\textwidth]{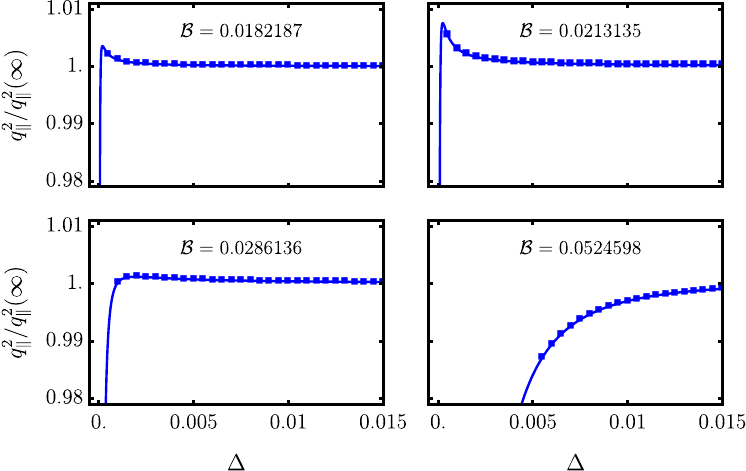}
    \caption{Discrete solutions for the order parameter (normalized by its asymptotic limit) as a function of $\Delta$, for fixed $\mathcal{B}$ (filled squares). The continuous line represents the envelope function defined by Eq.\eqref{eq:Sigmoid_parallel} before the conditions of Eq.\eqref{eq_qj} have been applied.}
    \label{fig:q3VsDelta}
\end{figure}

\begin{figure*}
    \centering
    \includegraphics[width=1\textwidth]{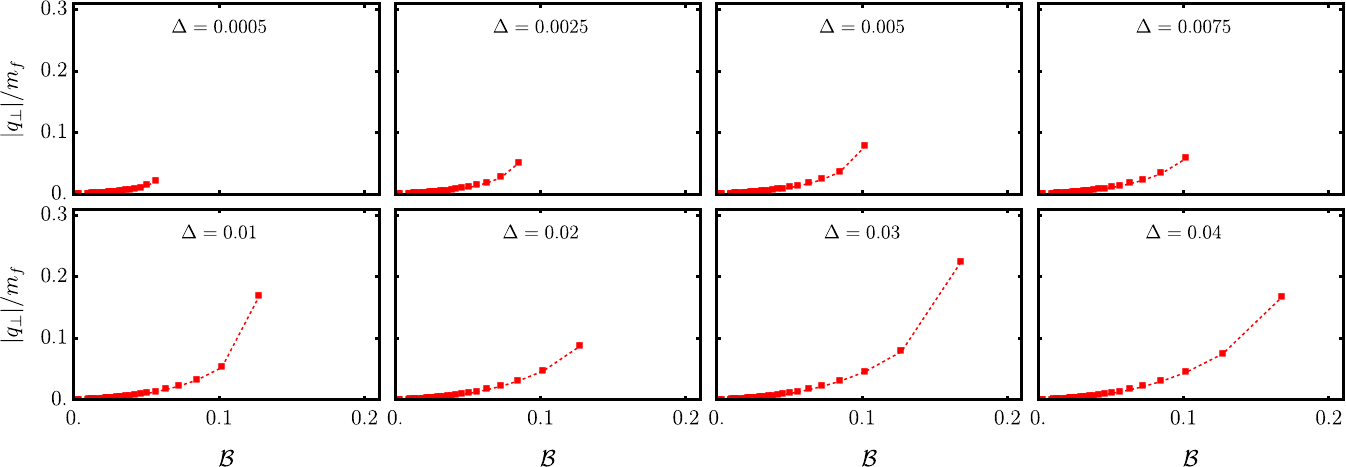}
    \caption{Non-trivial solutions of Eq.~\eqref{eq:det1} for the Case 2 as a function of $\mathcal{B}$ for $\Delta\in[0.1,0.04]$. The dashed line is the smooth envelope connecting those discrete non-trivial solutions.}
    \label{fig:qpVseB1}
\end{figure*}

\begin{figure}
    \centering
    \includegraphics[scale=0.94]{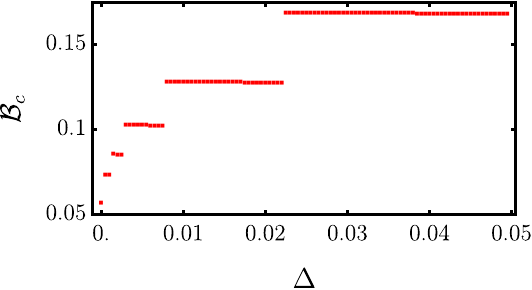}
    \caption{Critical magnetic field $\mathcal{B}_c$ of Fig.~\ref{fig:qpVseB1} as a function of $\Delta$ for the Case 2.}
    \label{fig:Bc_vs_Delta_high_1_qp}
\end{figure}

\begin{figure}
    \centering
    \includegraphics[width=0.5\textwidth]{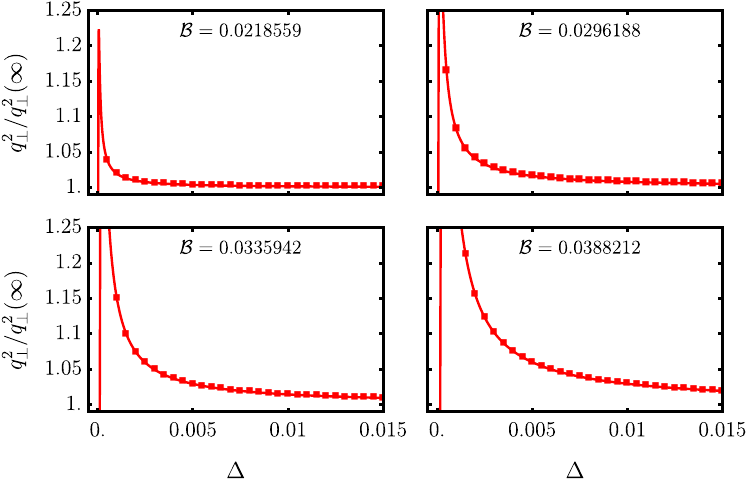}
    \caption{Discrete solutions for the order parameter (normalized by its asymptotic limit) as a function of $\Delta$, for fixed $\mathcal{B}$ (filled squares). The continuous line represents the envelope function defined by Eq.\eqref{eq:Sigmoid_perp} before the conditions of Eq.\eqref{eq_qj} have been applied.}
    \label{fig:qperpVsDelta}
\end{figure}
\subsection{Case 2: $q_1^2 = q_2^2 \equiv q_{\perp}^2$, with $q_3 =0$ }

In this case, the matrix Eq.~\eqref{eq_Mtilde_gen} reduces to
\bea
&&\left[\tilde{\mathcal{M}}[q_{\perp}]\right]^{jl} = \left(-\frac{1}{2}\delta^{jl}\left( \mathcal{M}^{11} + \mathcal{M}^{22} + \mathcal{M}^{12} + \mathcal{M}^{21} \right)\right.\nonumber\\
&&\left.+ e^2 \Delta_{B}
\left( \mathcal{M}^{jl11} + \mathcal{M}^{jl22} + \mathcal{M}^{jl12} + \mathcal{M}^{jl21}     \right)
\right) q_{\perp}^2\nonumber\\
&&\equiv \mathcal{C}_{\perp}^{jl} q_{\perp}^2.
\eea

For this case, the secular Eq.~\eqref{eq_secular} reduces to
\bea
{\rm{det}}\left( \Delta_B \mathcal{M} + q_{\perp}^2 \mathcal{C}_{\perp}   \right) = 0,
\eea
and repeating the same procedure as described in Case 1, we obtain
\bea
{\rm{det}}\left( \mathbf{1}_3 + q_{\perp}^2 \Delta_B^{-1} \mathcal{M}^{-1} \mathcal{C}_{\perp}   \right) = 0.
\label{eq_sec_case2}
\eea

Finally, just as in Case 1, we retain only second order powers of $q_{\perp}$ in Eq.~\eqref{eq_sec_case2} to arrive at the explicit algebraic solution
\bea
q_{\perp}^2 = -\frac{\Delta_B}{{\rm{tr}}\left( \mathcal{M}^{-1} \mathcal{C}_{\perp}  \right)}= \frac{\chi_{\perp}\Delta_B}{1 + \mathcal{K}_{\perp}\Delta_B},
\label{eq:Sigmoid_perp}
\eea
where we defined the parameters
\bea
\chi_{\perp} &=& \frac{2}{\left(\sum_{m,n=1,2}\mathcal{M}^{mn}\right)\,{\rm{tr}}\left( \mathcal{M}^{-1}\right)},\nonumber\\
\mathcal{K}_{\perp} &=& \frac{-2e^2\sum_{m,n=1,2}{\rm{tr}}\left( \mathcal{M}^{-1} \mathcal{\tilde{M}}^{(mn)}  \right)}{\left(\sum_{m,n=1,2}\mathcal{M}^{mn}\right)\,{\rm{tr}}\left( \mathcal{M}^{-1}\right)},
\eea
and the reduced matrices
\bea
\left[\mathcal{\tilde{M}}^{(mn)}\right]^{jl}
= \mathcal{\tilde{M}}^{jlmn}.
\eea

As already discussed in Eq.~\eqref{eq_qj}, the order parameter $q_j$ (for $j=1,2,3$) is a pure imaginary number.

Figure \ref{fig:qpVseB1} illustrates the non-trivial solutions for $q_\perp$ in Case 2, considering various values of $\Delta$. Similar to Case 1, these solutions are discrete and highly dependent on the magnetic field and noise parameter. Notably, the critical magnetic field in this scenario is lower compared to Case 1. Furthermore, Fig.~\ref{fig:Bc_vs_Delta_high_1_qp} demonstrates the behavior of the critical magnetic field $\mathcal{B}_c$, which exhibits a distinct functional form from that shown in Fig.~\ref{fig:Bc_vs_Delta_high_1}. The relationship between $q_\perp^2$ and $\Delta$, with an envelope given by Eq.~\eqref{eq:Sigmoid_perp}, is depicted in Fig.~\ref{fig:qperpVsDelta}, where discrete non-trivial solutions at particular values of $\Delta$ are permissible for a constant magnetic field. 

Here, we can also identify the value at which the sigmoidal Eq.~\eqref{eq:Sigmoid_perp} saturates as a function of the noise $\Delta_B$,
\bea
q_\perp^2(\infty)=\lim_{\Delta\to\infty}q_\perp^2=\frac{\chi_\perp}{\mathcal{K}_\perp}.
\eea

In the opposite limit, for very small values of $\Delta$, Eq.~\eqref{eq:Sigmoid_perp} shows that the order parameter follows an approximately linear trend with a slope defined by

\bea
\chi_{\perp} = \lim_{\Delta_B\rightarrow 0}\frac{\partial}{\partial \Delta_B} q_{\perp}^2.
\eea

Consequently, we can infer that the underlying physics in both cases is closely related, and the mechanism leading to the emergence of a non-trivial value of the order parameter maintains a consistent nature and interpretation, regardless of the specific values of $(q_1, q_2, q_3)$.


\subsection{Physical interpretation of the order parameter $Q_j$}
The physical interpretation of the order parameter components can be obtained from the functional representation Eq.~\eqref{eq_barZn}, {\it{before}} we integrate out the fermions, to obtain via saddle-point the mean expectation value

\begin{eqnarray}
Q_j = \ii e \Delta_B \langle \langle \bar{\psi}\gamma_j \psi\rangle\rangle_{\Delta} = - Q_j^*,
\end{eqnarray}

where here the double bracket stands for the statistical average over the classical noise in the background field, as well as the quantum expectation value of the corresponding observable, which clearly is a component of the vector current for the fermions, $j = 1,2,3$.

It is important to remark the physical effect that this order parameter, at the mean-field level, produces in the disorder-averaged fermion propagator, that is given by 
\begin{equation}
\ii S_{\text{F},\Delta}^{-1} (x - y) = \left( \ii\slashed{\partial} - e \slashed{A}_\text{BG} - m_f - e \slashed{q} \right)_x\delta^{(4)}(x-y), 
\end{equation}
where we defined $\slashed{q} = \gamma^{j} q_{j}$, for $q_j = Q_{j} - Q_{j}^{*}$ the order parameter. Clearly then, the differential equation for the propagator in the presence of the magnetic noise is given by
\begin{eqnarray}
\left( \ii\slashed{\partial} - e\slashed{A}_\text{BG} - m_f - e \slashed{q}  \right)_x S_{\text{F},\Delta}(x-y) = \ii \delta^{(4)}(x - y).
\label{eq_SDelta}
\end{eqnarray}

It is straightforward to verify that, if $S_\text{F} (x-y)$ represents the Schwinger propagator in the presence of the average background magnetic field, then the function
\begin{eqnarray}
S_{\text{F},\Delta}(x-y) = e^{- \ii e q\cdot(x-y)} S_\text{F}(x-y)
\label{eq_disphase}
\end{eqnarray}
is a solution to Eq.~\eqref{eq_SDelta}. Indeed, by direct substitution, we have
\begin{widetext}
\begin{eqnarray}
\left( \ii\slashed{\partial} - e\slashed{A}_\text{BG} - m_f - e \slashed{q}  \right)_x S_{\text{F},\Delta}(x-y)
&=& \left( \ii\slashed{\partial} - e\slashed{A}_\text{BG} - m_f - e \slashed{q}  \right)_x \left[ e^{- \ii e q\cdot(x-y)} S_F(x-y) \right]\nonumber\\
&=& e^{- \ii e q\cdot(x-y)} \left( \ii\slashed{\partial}  - e\slashed{A}_\text{BG} - m_f   \right)_x S_\text{F}(x-y) = \ii \delta^{(4)}(x - y),
\end{eqnarray}
\end{widetext}
where in the last step we applied the definition Eq.~\eqref{eq_Schwingernonoise} for the Schwinger propagator in the absence of noise.

Therefore, we conclude that at the level of the disorder-averaged propagator, Eq.~\eqref{eq_disphase} introduces an exponential damping effect given that the order parameter $q_j$ is a pure imaginary number.

\section{Summary and Conclusions}

In this work, we considered a system of QED fermions submitted to an external, classical magnetic field. In particular, we studied the effects of white noise in this magnetic field with respect to an average uniform value $\langle\mathbf{B}\rangle_{\Delta} = \hat{\mathbf{e}}_3 B$, as a function of the standard deviation $\Delta_B$, over the fermion propagator. As we discussed in the Introduction, this represents a statistical model for the actual scenario in heavy-ion collisions, where strong magnetic fields emerge for very short times within small spatial regions, whose size is of the order of the scattering cross-section. Since several such collisions occur at different points in space, the physical situation can be represented by a statistical ensemble, for different realizations of the magnetic field fluctuations, which are then described as a random variable.

We analyzed our model by applying the replica formalism, that led us to an effective action in terms of auxiliary bosonic fields. A mean field analysis of the corresponding effective action reveals that the magnetic noise effects can be captured by an order parameter, whose physical interpretation is the statistical ensemble average of the expectation value of the fermion vector current components. Therefore, non-trivial solutions where this order parameter acquires a non-zero value break the U(1) gauge symmetry in the system, as a consequence of the statistical noise in the background magnetic field. An interesting feature of such non-trivial solutions is that they exist only for certain discrete values of the average background magnetic field. Such discrete values can be identified to be in correspondence with the quantized Landau levels associated to the average background field. This feature is then consistent with the interpretation of the order parameter as the ensemble average of the fermion current. In addition, for a fixed value of the disorder strength characterized by $\Delta_B$, we find an upper critical value of the average background magnetic field $\mathcal{B}_c$, beyond  which the non-trivial solutions cease to exist in favour of a vanishing order parameter. This region of parameter space is then characterized by a dominance of the average background field over noise, whose effect then becomes negligible. In contrast, in the limit of very strong magnetic noise $\Delta_B\rightarrow\infty$, we observe that the order parameter asymptotically saturates to a constant finite value $q_{\parallel,\perp}^2(\infty)$ that depends on the field, but is independent of $\Delta_B$, as can be clearly seen from Eq.~\eqref{eq:Sigmoid_parallel} and Eq.~\eqref{eq:Sigmoid_perp}, respectively.

Remarkably, in the context of the fermion propagator, we showed that the order parameter, which is strictly imaginary, represents a finite screening length that leads to weak localization effects. Our present analysis is restricted to the fundamental level of the fermion propagator, but its consequences could manifest themselves in physical observables, such as effective collision rates for certain processes.

\acknowledgements{ J.D.C.-Y. and E.M. acknowledge financial support from ANID PIA Anillo ACT/192023. E.M. also acknowledges financial support from Fondecyt 1230440. M. L. acknowledges support from ANID/CONICYT FONDECYT Regular (Chile) under grants No. 1200483, 1190192 and 1220035. 

\bibliography{References.bib}

\newpage
\appendix
\begin{widetext}
\section{Traces involved in the definition of the matrices $\mathcal{M}$}
In this Appendix, we provide an explicit example of the method used to calculate the traces of products of operators involved in the definition of the matrices $\mathcal{M}^{ij}$ and $\mathcal{M}^{ijkl}$.

For definiteness, let us consider the following expression
\bea
\mathcal{M}^{jlmn}(k)\equiv\text{Tr}\left[S_\text{F}(k)\gamma^j S_\text{F}(k)\gamma^l S_\text{F}(k)\gamma^m S_\text{F}(k)\gamma^n\right].
\eea
To calculate the traces over Dirac matrices, note that the propagator product yields to several terms for $\mathcal{M}^{jlmn}$, so that we can define:
\bea
\mathcal{M}^{jlmn}(k)&=&\sum_{a=1}^3\sum_{b=1}^3\sum_{c=1}^3\sum_{d=1}^3\mathcal{A}_a(k)\mathcal{A}_b(k)\mathcal{A}_c(k)\mathcal{A}_d(k)\text{Tr}\left[s_a(k)\gamma^j s_b(k)\gamma^l s_c(k)\gamma^m s_d(k)\gamma^n\right]\nn\\
&\equiv& \sum_{a=1}^3\sum_{b=1}^3\sum_{c=1}^3\sum_{d=1}^3\mathcal{A}_a(k)\mathcal{A}_b(k)\mathcal{A}_c(k)\mathcal{A}_d(k)\mathcal{M}^{jlmn}_{abcd}(k)\nn\\
\eea
where
\bea
s_1(k)&\equiv&\slashed{k}_\parallel+m\nn\\
s_2(k)&\equiv&\ii\gamma^1\gamma^2\left(\slashed{k}_\parallel+m\right)\nn\\
s_3(k)&\equiv&\slashed{k}_\perp.
\eea

Note that for the cyclic property of the trace the elements satisfy:
\bea
\mathcal{M}^{jlmn}_{abcd}=\mathcal{M}^{njlm}_{abcd}=\mathcal{M}^{mnjl}_{abcd}=\mathcal{M}^{lmnj}_{abcd},
\label{Eq:propiedadciclicageneral}
\eea
ans therefore, just a few traces needs to be computed explicitly, so that the whole expression can be reached by adding terms with the convenient indexes. The needed elements $\mathcal{M}^{jlmn}_{\alpha\beta\sigma\rho}(k)$ can be straightforward computed:

\bea
\mathcal{M}^{jlmn}_{1111}&=&4\Big[8\kp^j\kp^l\kp^m\kp^n+2(m^2-\kp^2)\left(g^{jn}\kp^l\kp^m+g^{mn}\kp^j\kp^l+g^{lm}\kp^j\kp^n+g^{jl}\kp^m\kp^n\right)\nn\\
&+&(m^2-\kp^2)^2\left(g^{jn}g^{lm}-g^{jm}g^{ln}\right)+(m^4+\kp^4)g^{jl}g^{mn}\Big]
\eea

\bea
\mathcal{M}^{jlmn}_{3333}&=&4\Bigg[8\kt^j\kt^l\kt^m\kt^n+\kt^4\left(g^{jn}g^{mn}-g^{jm}g^{ln}+g^{jn}g^{lm}\right)\nn\\
&-&2\kt^2\left(\kt^j\kt^lg^{mn}+\kt^l\kt^mg^{jn}+\kt^j\kt^ng^{lm}+\kt^m\kt^ng^{jl}\right)\Bigg]\nn\\
\eea

\bea
\mathcal{M}^{jlmn}_{1112}&=&4\ii(\kp^2-m^2)\epsilon_{ab}^\perp\Bigg[2\left(\gt^{an}\gt^{bm}\kp^j\kp^l+\gt^{an}\gt^{bj}\kp^l\kp^m+\gt^{al}\gt^{bm}\kp^k\kp^n+\gt^{aj}\gt^{bl}\kp^m\kp^n\right)\nn\\
&+&(\kp^2-m^2)\left(\gt^{am}\gt^{bn}g^{jl}+\gt^{an}\gt^{bl}g^{jm}+
\gt^{am}\gt^{bl}g^{jn}+\gt^{aj}\gt^{bn}g^{lm}+\gt^{aj}\gt^{bm}g^{ln}+\gt^{al}\gt^{bj}g^{mn}\right)\Bigg],
\eea
where $\epsilon_{ab}^\perp$ is given by $\epsilon^\perp_{12}=-\epsilon^\perp_{21}=1$. 

The element $\mathcal{M}^{jlmn}_{1122}$ is conveniently splitted in two pieces, i.e., 
\bea
\mathcal{M}^{jlmn}_{1122}&=&\text{Tr}\left[(\slashed{k}_\parallel+m)\gamma^j(\slashed{k}_\parallel+m)\gamma^l\ii\gamma^1\gamma^2(\slashed{k}_\parallel+m)\gamma^m\ii\gamma^1\gamma^2(\slashed{k}_\parallel+m)\gamma^n\right]\nn\\
&=&\text{Tr}\left[(\slashed{k}_\parallel+m)\gamma^j(\slashed{k}_\parallel+m)\gamma^l(\slashed{k}_\parallel+m)\gamma^m_\parallel(\slashed{k}_\parallel+m)\gamma^n\right]-\text{Tr}\left[(\slashed{k}_\parallel+m)\gamma^j(\slashed{k}_\parallel+m)\gamma^l(\slashed{k}_\parallel+m)\gamma^m_\perp(\slashed{k}_\parallel+m)\gamma^n\right]\nn\\
&\equiv&\mathcal{M}^{jlmn}_{1122}(\text{a})+\mathcal{M}^{jlmn}_{1122}(\text{b}),
\eea
where we have used:
\bea
\gamma^1\gamma^2\gamma^\mu\gamma^1\gamma^2=-\gamma^\mu_\parallel+\gamma^\mu_\perp,
\label{Eq:identityGamma1Gamma2}
\eea
and we defined:
\begin{subequations}
  \bea
\mathcal{M}^{jlmn}_{1122}(\text{a})=\text{Tr}\left[(\slashed{k}_\parallel+m)\gamma^j(\slashed{k}_\parallel+m)\gamma^l(\slashed{k}_\parallel+m)\gamma^m_\parallel(\slashed{k}_\parallel+m)\gamma^n\right],
\eea
and
\bea
\mathcal{M}^{jlmn}_{1122}(\text{b})=-\text{Tr}\left[(\slashed{k}_\parallel+m)\gamma^j(\slashed{k}_\parallel+m)\gamma^l(\slashed{k}_\parallel+m)\gamma^m_\perp(\slashed{k}_\parallel+m)\gamma^n\right],
\eea
\end{subequations}
so that
\begin{subequations}
   \bea
\mathcal{M}^{jlmn}_{1122}(\text{a})&=&4\Bigg[8\kp^j\kp^l\kp^m\kp^n+\left(\kp^2-m^2\right)^2\left(\gp^{lm}g^{jn}-\gp^{jm}g^{ln}+\gp^{mn}g^{jl}\right)\nn\\
&+&2\left(m^2-\kp^2\right)\left(\kp^j\kp^l\gp^{mn}+\kp^l\kp^mg^{jn}+\kp^j\kp^n\gp^{lm}+\kp^m\kp^ng^{jl}\right)\Bigg],
\eea
and
    \bea
\mathcal{M}^{jlmn}_{1122}(\text{b})&=&4(\kp^2-m^2)\left[2\kp^j\left(\kp^l\gt^{mn}+\kp^n\gt^{lm}\right)+(\kp^2-m^2)\left(\gt^{jm}g^{lm}-\gt^{lm}g^{jn}-\gt^{mn}g^{jl}\right)\right].
\eea
\end{subequations}

\bea
\mathcal{M}^{jlmn}_{1133}&=&4\Big\{4\kp^j\kp^n\kt^l\kt^m+4\kp^j\kp^l\kt^m\kt^n-2\kt^2\kp^j\left(\kp^lg^{mn}-\kp^mg^{ln}+\kp^ng^{lm}\right)\nn\\
&+&(\kp^2-m^2)\left[\kt^2\left(g^{jl}g^{mn}-g^{jm}g^{ln}+g^{lm}g^{jn}\right)+2\kt^m\left(\kt^jg^{ln}-\kt^lg^{jn}-\kt^ng^{lj}\right)\right]\Big\}
\eea

By following the same procedure, the term $\mathcal{M}^{jlmn}_{2233}$ is spplited into:
\begin{subequations}
  \bea
\mathcal{M}^{jlmn}_{2233}(\text{a})&=&\text{Tr}\left[(\slashed{k}_\parallel+m)\gamma^j_\parallel(\slashed{k}_\parallel+m)\gamma^l\slashed{k}_\perp\gamma^m\slashed{k}_\perp\gamma^n\right]\nn\\
&=&4\Big\{(m^2-\kp^2)\left[2\kt^j\kt^m\gp^{jn}+2\kt^m\kt^n\gp^{jl}+\kt^2\left(\gp^{jm}g^{ln}-\gp^{jn}g^{lm}-\gp^{jl}g^{mn}\right)\right]\nn\\
&+&2\kp^j\left[\kt^2\left(\kp^mg^{ln}-\kp^ng^{lm}\right)+\kp^l(2\kt^m\kt^n-\kt^2g^{mn})+2\kp^n\kt^l\kt^m\right]\Big\}
\eea
and
\bea
\mathcal{M}^{jlmn}_{2233}(\text{b})&=&-\text{Tr}\left[(\slashed{k}_\parallel+m)\gamma^j_\perp(\slashed{k}_\parallel+m)\gamma^l\slashed{k}_\perp\gamma^m\slashed{k}_\perp\gamma^n\right]\nn\\
&=&4(m^2-\kp^2)\left[2\kt^m\left(\kt^jg^{ln}-\kt^n\gt^{jl}-\kt^l\gt^{jn}\right)+\kt^2\left(\gt^{jl}g^{mn}-\gt^{jm}g^{ln}+\gt^{jn}g^{lm}\right)\right].
\eea
\end{subequations}

\end{widetext}

\section{Momentum integrals}
In the calculation of the matrix coefficients, such as the example provided in Appendix B, we need to obtain momentum integrals of the general form,
\begin{eqnarray}
I^{\alpha,\beta,\gamma}_{\delta,\sigma} = \int \frac{d^4 k}{(2\pi)^4}
\left[\mathcal{A}_1(k) \right]^{\alpha}\left[\mathcal{A}_2(k) \right]^{\beta}
\left[\mathcal{A}_3(k) \right]^{\gamma}\left[ k_{\perp}^2\right]^{\delta}\left[k_{\parallel}^2 \right]^{\sigma}\nonumber\\
\label{eq_Ialphabeta}
\end{eqnarray}
Here, we recall the representation we obtained in Ref.~\cite{Replicas_1} for the function
\begin{eqnarray}
\mathcal{A}_1 (k) &=& \frac{\ii\,e^{-\frac{k_{\perp}^2}{eB}}}{2 e B} e^{-\frac{\ii\pi}{2 e B}\left( k_{\parallel}^2 - m_f^2 + \ii\epsilon  \right)}\Gamma\left(-\frac{k_{\parallel}^2 - m_f^2 + \ii\epsilon}{2 e B}  \right)\nonumber\\ &&\times U\left( -\frac{k_{\parallel}^2 - m_f^2 + \ii\epsilon}{2 e B},0,\frac{2k_{\perp}^2}{e B} \right),
\end{eqnarray}
where $\Gamma(z)$ is the Gamma function, while $U(a,b,z)$ is Tricomi's Confluent Hypergeometric function.

As a first step, let us perform a Wick rotation to recover the Euclidean metric, $k_{0}\rightarrow \ii k_{0}$, which implies $d^4 k \rightarrow \ii d^4 k$, and $k_{\parallel}^2\rightarrow - k_{\parallel}^2$ (here we avoid adding further sub-indexes to keep the notation simple). Moreover, let´s define the following auxiliary variables
\begin{eqnarray}
a &=& \frac{k_{\parallel}^2 + m_f^2}{2 e B}\nonumber\\
z &=& \frac{2k_{\perp}^2}{e B}
\end{eqnarray}
In terms of these new variables, we can write
\begin{eqnarray}
\mathcal{A}_1 (a,z) = \frac{\ii e^{-z/2}}{2 e B}
e^{\ii \pi a} \Gamma(a) U(a,0,z),
\end{eqnarray}
and the integration measure $\ii d^4 k = \ii d^2 k_{\perp} d^2 k_{\parallel}$, with
\begin{eqnarray}
d^{2}k_{\perp} &=& \pi d\left(k_{\perp}^2\right) = \frac{\pi e B}{2} dz\nonumber\\
d^{2}k_{\parallel}  &=& \pi d\left(k_{\parallel}^2\right) = 2\pi e B da
\end{eqnarray}

Therefore, Eq.~\eqref{eq_Ialphabeta} reduces to the expression
\begin{eqnarray}
I^{\alpha,\beta,\gamma}_{\delta,\sigma}
&=& \frac{\pi^2 (eB)^{2 + \delta}}{ 2^{\delta} (2\pi)^4}
\int_{\frac{m_f^2}{2 e B}}^{\infty} da \left(  2 e B a + m_f^2\right)^{\sigma}\\
&&\times\int_{0}^{\infty}dz\,z^{\delta} \left[\mathcal{A}_1 (a,z) \right]^{\alpha}\left[\mathcal{A}_2 (a,z) \right]^{\beta}
\left[\mathcal{A}_3 (a,z) \right]^{\gamma}\nonumber
\end{eqnarray}

In order to calculate the integrals in this last form, we shall apply the identity
\begin{eqnarray}
\Gamma(a) U(a,\epsilon,z) &&= \frac{1}{a} M(a,\epsilon,z) + \Gamma(-1+\epsilon)M(1+a,2,z)\nonumber\\
&&\sim \frac{1}{a}M(a,\epsilon,z) + \left(  \gamma_e - 1\right) z M(1+a,2,z)\nonumber\\
\end{eqnarray}
where $M(a,b,z)$ represents Kummer's Confluent Hypergeometric function, and $\Gamma(z)$ is the Gamma function. In the second line, we have removed the $1/\epsilon$ divergence of the $\Gamma(-1+\epsilon)$ function. Finally, we regularize by subtracting the divergent term $M(a,\epsilon,z)$ as $\epsilon\rightarrow 0$, to arrive at the prescription
\begin{eqnarray}
\Gamma(a) U(a,\epsilon,z) &\simeq& (\gamma_e - 1)z M(1+a,2,z)\\
&=& (\gamma_e - 1) z \left( 1 + \frac{1+ a}{2}z\right) + O(z^3)\nonumber
\end{eqnarray}

From this expansion, we obtain
\begin{eqnarray}
\mathcal{A}_1(a,z) \simeq (\gamma_e - 1)\frac{\ii e^{-z/2}}{2 e B}e^{\ii\pi a}  z \left( 1 + \frac{1+ a}{2}z\right).
\label{eq_A1expand}
\end{eqnarray}

The other functions are expressed by the following expansions at the same order
\begin{eqnarray}
\mathcal{A}_2(a,z) &=& 2 \ii \frac{\partial\mathcal{A}_1(a,z)}{\partial z}\nonumber\\
&=& 2 (\ii)^2 (\gamma_e - 1) \frac{e^{\ii\pi a}}{2 e B} e^{-z/2}\left[1 + \left(a + \frac{1}{2}\right)z\right.\nonumber\\
&&\left.-\frac{1+a}{4}z^2\right]
\end{eqnarray}
and
\begin{eqnarray}
\mathcal{A}_3(a,z) &=& \mathcal{A}_1(a,z) + (2 \ii)^2 \frac{\partial^2\mathcal{A}_1}{\partial z^2}\\
&=& -4\,\ii(\gamma_e - 1) \frac{e^{\ii\pi a}}{2 e B} e^{-z/2}\left[ a - (1 + a)z  \right]\nonumber 
\end{eqnarray}

Inserting into the integral expression, we obtain
\begin{eqnarray}
I^{\alpha,\beta,\gamma}_{\delta,\sigma}
&=& \frac{\pi^2 (eB)^{2 + \delta}}{ 2^{\delta} (2\pi)^4}(2 \ii)^{\beta} (-4)^{\gamma}\left[\ii\frac{(\gamma_e - 1)}{2 e B}\right]^{\alpha + \beta + \gamma}\nonumber\\
&&\times\int_{\frac{m_f^2}{2 e B}}^{\infty}  da\,e^{\ii\pi(\alpha + \beta + \gamma)a} \left(  2 e B a + m_f^2\right)^{\sigma}\\
&&\times\int_{0}^{\infty}dz\,e^{-\frac{(\alpha + \beta + \gamma)z}{2}} z^{\delta+\alpha} \left[1 + \frac{1+a}{2}z\right]^{\alpha}\nonumber\\
&&\times\left[1 + \left( a + \frac{1}{2} \right)z  - \frac{1+a}{4}z^2\right]^{\beta}\nonumber\\
&&\times\left[a - (1+a)z \right]^{\gamma}\nonumber
\label{eq_Ialphabetagamma}
\end{eqnarray}

Now, let us expand the binomials and trinomials in the integrand as follows
\begin{eqnarray}
\left[1 + \frac{1+a}{2} z \right]^{\alpha}
= \sum_{m=0}^{\alpha}\frac{\alpha!}{m! (\alpha-m)!}\left( \frac{1+a}{2}  \right)^{m}z^m
\end{eqnarray}
\begin{eqnarray}
&&\left[1 + \left(a + \frac{1}{2}\right)z - \frac{1+a}{4}z^2  \right]^{\beta}= \sum_{q=0}^{\beta}\sum_{j=0}^{q}\frac{\beta!}{j!\left( \beta - q  \right)!\left( q-j \right)!}\nonumber\\
&&\times\left(a + \frac{1}{2}\right)^j \left(-\frac{(1+a)}{4}  \right)^{q-j} z^{2q-j}
\end{eqnarray}
\begin{eqnarray}
\left[a - (1 + a)z \right]^{\gamma} = \sum_{n=0}^{\gamma}\frac{\gamma!}{n! (\gamma -n)!} a^{\gamma - n} \left(-(1+a)\right)^{n}z^n.\nn\\
\end{eqnarray}

Inserting these expansions into Eq.\eqref{eq_Ialphabetagamma}, we obtain
\begin{widetext}
\begin{eqnarray}
I^{\alpha,\beta,\gamma}_{\delta,\sigma}
&=& \frac{\pi^2 (eB)^{ \delta + \sigma}(2\ii)^{\beta}(-4)^{\gamma}}{2^{ \delta - \sigma} (2\pi)^4}\left[\ii\frac{(\gamma_e - 1)}{2 e B}\right]^{\alpha + \beta + \gamma}\sum_{m=0}^{\alpha}
\frac{\alpha!2^{-m}}{m! (\alpha - m)!}\sum_{n=0}^{\gamma}
\frac{\gamma!}{n! (\gamma - n)!}\sum_{q=0}^{\beta}\sum_{j=0}^{q}\frac{\beta!4^{j-q}}{j!(\beta - q)! (q - j)!}\nonumber\\
&&\times\int_{\frac{m_f^2}{2 e B}}^{\infty}  da\,e^{\ii\pi(\alpha + \beta + \gamma)a} \left(   a + \frac{m_f^2}{2 e B}\right)^{\sigma}
a^{\gamma - n}\left(-(1 + a)  \right)^{n + q - j}(1+a)^m \left(  a + \frac{1}{2}\right)^{j}
\\
&&\times\int_{0}^{\infty}dz\,e^{-\frac{(\alpha + \beta + \gamma)z}{2}} z^{\delta+\alpha + n + m + 2q - j }\nonumber
\label{eq_Ial2}
\end{eqnarray}

We further expand the binomials in the variable a, as follows
\begin{eqnarray}
\left(   a + \frac{m_f^2}{2 e B}\right)^{\sigma} = \sum_{k=0}^{\sigma}\frac{\sigma!}{k!(\sigma - k)!}\left( \frac{m_f^2}{2 e B} \right)^{\sigma - k}a^k
\end{eqnarray}
\begin{eqnarray}
\left(-(1 + a)  \right)^{n + q - j} = (-1)^{n + q - j}\sum_{l=0}^{n + q - j}\frac{(n + q - j)!}{l!(n + q - j-l)!}a^l
\end{eqnarray}
\begin{eqnarray}
\left(  a + \frac{1}{2}\right)^{j} = \sum_{h=0}^{j}\frac{j!}{h! (j-h)!}2^{-(j-h)}a^h
\end{eqnarray}

Inserting these expansions into Eq.~\eqref{eq_Ial2}, we obtain
\begin{eqnarray}
I^{\alpha,\beta,\gamma}_{\delta,\sigma}
&=& \frac{\pi^2 (eB)^{2 + \delta + \sigma}(2\ii)^{\beta}(-4)^{\gamma}}{2^{ \delta - \sigma} (2\pi)^4}\left[\ii\frac{(\gamma_e - 1)}{2 e B}\right]^{\alpha + \beta + \gamma}\sum_{m=0}^{\alpha}
\frac{\alpha!2^{-m}}{m! (\alpha - m)!}\sum_{n=0}^{\gamma}
\frac{\gamma!}{n! (\gamma - n)!}\sum_{q=0}^{\beta}\sum_{j=0}^{q}\frac{\beta!4^{j-q}}{j!(\beta - q)! (q - j)!}\nonumber\\
&&\times\sum_{k=0}^{\sigma}\frac{\sigma!}{k!(\sigma - k)!}\left( \frac{m_f^2}{2 e B} \right)^{\sigma - k}
(-1)^{n + q - j}\sum_{l=0}^{n + q - j}\frac{(n + q - j)!}{l!(n + q - j-l)!}
\sum_{h=0}^{j}\frac{j!}{h! (j-h)!}2^{-(j-h)}\nonumber\\
&&\times\left[ \int_{\frac{m_f^2}{2 e B}}^{\infty} da\, e^{\ii\pi (\alpha + \beta + \gamma + \ii\epsilon)a}\,a^{\gamma - n + k + l + h} \right]
\left[ \int_0^{\infty} dz\, e^{-\frac{(\alpha+\beta+\gamma)z}{2}}
z^{\delta+\alpha + n + m + 2q - j} \right]
\end{eqnarray}

It is now trivial to show the identities
\begin{eqnarray}
\int_0^{\infty} dz\, e^{-\frac{(\alpha+\beta+\gamma)z}{2}}
z^{\delta+\alpha + n + m + 2q - j} = \left( \frac{2}{\alpha + \beta + \gamma} \right)^{\alpha + \delta + n + m + 2q-j + 1}\cdot( \alpha + \delta + n + m + 2q-j  )!
\end{eqnarray}
and
\begin{eqnarray}
\int_{\frac{m_f^2}{2 e B}}^{\infty} da\, e^{\ii\pi (\alpha + \beta + \gamma + \ii\epsilon)a} = \frac{\ii}{\pi\left(  \alpha + \beta + \gamma\right)} e^{\ii\pi\left(\alpha + \beta + \gamma\right)\frac{m_f^2}{2 e B}}
\end{eqnarray}
\begin{eqnarray}
\int_{\frac{m_f^2}{2 e B}}^{\infty} da\, e^{\ii\pi (\alpha + \beta + \gamma + \ii\epsilon)a}\,a^{n} = \frac{\ii^{n+1}}{\left[\pi\left(  \alpha + \beta + \gamma\right)\right]^{n+1}} \Gamma\left( n+1,-\ii\pi(\alpha + \beta + \gamma)  \frac{m_f^2}{2 e B}\right).
\end{eqnarray}

Inserting these identities, we finally obtain
\begin{eqnarray}
I^{\alpha,\beta,\gamma}_{\delta,\sigma}
&=& \frac{\pi^2 (eB)^{2 + \delta + \sigma}(2\ii)^{\beta}(-4)^{\gamma}}{2^{ \delta - \sigma} (2\pi)^4}\left[\ii\frac{(\gamma_e - 1)}{2 e B}\right]^{\alpha + \beta + \gamma}\sum_{m=0}^{\alpha}
\frac{\alpha!2^{-m}}{m! (\alpha - m)!}\sum_{n=0}^{\gamma}
\frac{\gamma!}{n! (\gamma - n)!}\sum_{q=0}^{\beta}\sum_{j=0}^{q}\frac{\beta!4^{j-q}}{j!(\beta - q)! (q - j)!}\nonumber\\
&&\times\sum_{k=0}^{\sigma}\frac{\sigma!}{k!(\sigma - k)!}\left( \frac{m_f^2}{2 e B} \right)^{\sigma - k}
(-1)^{n + q - j}\sum_{l=0}^{n + q - j}\frac{(n + q - j)!}{l!(n + q - j-l)!}
\sum_{h=0}^{j}\frac{j!}{h! (j-h)!}2^{-(j-h)}\nonumber\\
&&\times\left[ \frac{\ii^{n+1}}{\left[\pi\left(  \alpha + \beta + \gamma\right)\right]^{\gamma - n + k + l + h +1}} \Gamma\left( \gamma -n + k + l + h +1,-\ii\pi(\alpha + \beta + \gamma)  \frac{m_f^2}{2 e B}\right)\right]\nonumber\\
&&\times\left( \frac{2}{\alpha + \beta + \gamma} \right)^{\alpha + \delta + m + n + 2q-j + 1}\cdot( \alpha + \delta + n + m + 2q-j  )!
\end{eqnarray}
\end{widetext}

\end{document}